\documentclass[prl,twocolumn,preprintnumbers,nofootinbib,showpacs,amsmath,amssymb,superscriptaddress]{revtex4-1}
\usepackage{psfrag,graphicx}

\usepackage[
      colorlinks=true,
      linkcolor=blue,
      urlcolor=blue,
      filecolor=black,
      citecolor=blue,
            ]{hyperref}
\usepackage{color}
\usepackage{amsfonts}
\usepackage{amssymb}
\usepackage{epstopdf}
\usepackage{dsfont}
\usepackage{epsfig}

\begin{document}
\title{ Holographic model for antiferromagnetic quantum phase transition induced by magnetic field}
\author{Rong-Gen Cai}
\email{cairg@itp.ac.cn}

\author{Run-Qiu Yang}
\email{aqiu@itp.ac.cn}
\affiliation{State Key Laboratory of Theoretical Physics,Institute of Theoretical Physics,\\
 Chinese Academy of Sciences,Beijing 100190, China.}

\author{F.V. Kusmartsev}
\email{F.Kusmartsev@lboro.ac.uk}
\affiliation{Department of Physics, Loughborough University, Loughborough, Leicestershire, LE11 3TU, United Kingdom}

\begin{abstract}
We  propose a  gravity dual of antiferromagnetic quantum phase transition (QPT) induced by magnetic field and  study the critical behavior around  the quantum critical point (QCP).
It turns out that the boundary critical theory is a strong coupling theory with dynamic exponent $z=2$ and that the hyperscaling law is violated and logarithmic corrections appear near the QCP. Some novel scaling relations are  predicated, which can be tested by experiment data in future. We also make some comparison with experimental data on
low-dimensional magnets BiCoPO$_5$ and  pyrochlores Er$_{2-2x}$Y$_{2x}$Ti$_2$O$_7$.

\end{abstract}
\maketitle

\noindent
\noindent { \it Introduction.}
--Quantum phase transition (QPT) and the behavior of quantum systems in the vicinity of the corresponding quantum critical point (QCP) have attracted a lot of attention both in theory and  experiment sides recently~\cite{S. Sachdev,Hertz.76,Sachdev_book}. In contrast to their classical counterparts  induced  by thermal fluctuations arising at finite temperature $T>0$, QPTs  happen at zero temperature and  are governed by  quantum fluctuations associated with the Heisenberg uncertainty and are driven by a certain  control parameter rather than temperature, e.g., composition, magnetic field or pressure, etc. In condensed matter physics, such a quantum criticality is considered to play an important role in some interesting phenomena~\cite{Gegenwart.08,Lohneysen.07}.

One of intensively discussed QPTs is ordered-disordered QPT in antiferromagnetic materials induced by magnetic field (see for example, Refs.~\cite{G.Chaboussant,W.Shiramura,Oosawa,Ch.03,Nikuni}),
especially in the heavy-fermion systems, since they can be tuned continuously from an antiferromagnetic (AF) state to a paramagnetic (PM) metallic state by varying a single parameter~\cite{Gegenwart.08}. In these materials, QPT naturally belongs to the phenomenon involving strongly correlated many-body systems~\cite{Si.01,Senthil.04,Tanaka}.  However, the complete theoretical descriptions valid in all the energy (or temperature) region are still lacking. In order to study and characterize strongly coupled quantum critical systems, some new methods are called for.

Thanks to the feature of the weak/strong duality,  the AdS/CFT correspondence  provides a powerful  approach to study such strongly coupled systems.
This duality relates a weak coupling gravitational theory in a $(d+1)$-dimensional asymptotically anti-de Sitter (AdS) space-time to a $d$-dimensional strong coupling conformal field theory (CFT) in the AdS boundary~\cite{Maldacena:1997re,Gubser:1998bc,Witten:1998qj}.
In recent years, we have indeed witnessed that the duality has been extensively applied into condensed matter physics and some significant progresses have been made~\cite{Hartnoll:2008vx,Lee:2008xf,Liu:2009dm,Cubrovic:2009ye}.  In Ref.~\cite{Cai:2014oca} we realized the
ferromagnetic/paramagnetic phase transition in a holographic setup,  and in Ref.~\cite{Cai:2014jta} the holographic antiferromagnetic/paramagnetic phase transition was studied. We showed that the antiferromagnetic transition temperature $T_N$ is indeed suppressed by an external magnetic field and tends to zero when the magnetic field reaches  its critical value $B_c$.  In this way  the antiferromagnetic QPT induced  by magnetic field is realized.  However, it was shown that  the model proposed in Ref.~\cite{Cai:2014oca} contains a vector ghost, very recently, a modified model  was proposed~\cite{Cai:2015bsa}, which is shown not only ghost free, but also causal well-defined, while it keeps  the main results in the original model qualitatively. Here we will elaborate in some detail this QPT and study the corresponding critical properties in this new model.

\noindent {\it Holographic model.}
--In order to describe the spontaneous staggered magnetization which breaks the time reversal symmetry, we introduce two real antisymmetric tensor fields coupled with U(1) Maxwell field strength~\cite{Cai:2014jta}. Based on the discussions in Refs.~\cite{Cai:2014jta,Cai:2015bsa},  we take the bulk action as follows,
\begin{equation}\label{action1}
S=\frac1{2\kappa^2}\int d^4x\sqrt{-g}[R+\frac{6}{L^2}-F^{\mu\nu}F_{\mu\nu}-\lambda^2(L_1+L_2+L_{12})],
\end{equation}
where
\begin{equation}\label{action2}
\begin{split}
&L_{12}=\frac k2M^{(1)\mu\nu}M^{(2)}_{\mu\nu},\\
&L_{(a)}=\frac1{12}(dM^{(a)})^{\mu\nu\tau}(dM^{(a)})_{\mu\nu\tau}+\frac{m^2}4 M^{(a)\mu\nu}M^{(a)}_{\mu\nu}\\
&~~~~~~~~~+\frac12M^{(a)\mu\nu}F_{\mu\nu}+J V(M^{(a)}_{\mu\nu}),\\
&~V(M^{(a)}_{\mu\nu})=({^*M^{(a)}}_{\mu\nu}M^{(a)\mu\nu})^2, ~a=1,2.
\end{split}
\end{equation}
Here $^*$ is the Hodge star dual operator and $dM$ denotes the exterior derivative of $M$. $L$ is the radius of AdS space, $2\kappa^2=16\pi G$ with $G$  the Newtonian gravitational  constant,  $k$, $m^2$ and $J$ are all model parameters with $J<0$,  $\lambda^2$ characterizes the back reaction of the two polarization fields $M^{(a)}_{\mu\nu}$ to the background geometry, and $L_{12}$ describes the interaction between two polarization fields.  Note that by rescaling the polarization fields  and the parameter $J$, $\lambda^2$ can also be viewed as the coupling strength between the polarization fields
and the background Maxwell field.  In the AdS/CFT duality,  the model parameters  $m$ and $k$  are related to the dual operator dimension in the boundary field theory,  $J$ is related to the self-coupling coefficient of the magnetic moment  and $k$ describes the interaction
between two kinds of magnetic moments. The reason to introduce two antisymmetric fields for describing the anti-ferromagnetism was elaborated in Ref.~\cite{Cai:2014jta}. Note that the form of $V(M^{(a)}_{\mu\nu})$ is not unique, we choose this form as it can lead to spontaneous symmetry breaking  (see  Fig.1 in Ref.~\cite{Cai:2015bsa}) and to simplify  the equations of motion of the model.  Compared with the original model for antiferromagnetism in  Ref.~\cite{Cai:2014jta},  the key change is to replace the covariant derivative of  the polarization field $M$
by the exterior derivative. This change can avoid the problems such as ghost and causal violation, while keep the significant results in the original model qualitatively and  in addition this model has a potential  origin in string/M theory~\cite{Cai:2015bsa}.

%
%
In the probe limit of $\lambda\rightarrow0$, we can neglect the back reaction of the two polarization fields on the background geometry. The background we will consider is a dyonic Reissner-Nordstr\"om-AdS black brane solution of
the Einstein-Maxwell theory with a negative cosmological constant, and the metric reads~\cite{Cai:1996eg}
\begin{equation}\label{geom}
\begin{split}
  ds^2=r^2(-f(r)dt^2+dx^2+dy^2)+\frac{dr^2}{r^2f(r)},\\
   f(r)=1-\frac{1+\mu^2+B^2}{r^3}+\frac{\mu^2+B^2}{r^4}.
\end{split}
\end{equation}
Here both the black brane horizon $r_h$ and AdS radius $L$ have been set to be unity. The temperature of the black
brane is
\begin{equation}\label{Tem1}
T=(3-\mu^2-B^2)/4\pi.
\end{equation}
For the solution (\ref{geom}), the corresponding gauge potential is $ A_\mu=\mu(1-1/r)dt+Bx dy$. Here $\mu$ is the chemical potential and $B$ can be viewed as the external magnetic field of the dual boundary field theory.

We consider a self-consistent ansatz for the tensor fields with nonvanishing  components $M^{(a)}_{tr}, M^{(a)}_{xy}$ (a=1,2) and define
\begin{equation}\label{ab1}
\alpha=(M^{(1)}_{xy}+M^{(2)}_{xy})/2,~~\beta=(M^{(1)}_{xy}-M^{(2)}_{xy})/2.
\end{equation}
By this definition, the antiferromagnetic order parameter, i.e., the staggered magnetization, can be expressed as~\cite{Cai:2014jta,Cai:2015bsa}
\begin{equation}\label{eqN1}
N^\dagger/\lambda^2=-\int dr\beta/r^2.
\end{equation}
Then the antiferromagnetism phase corresponds to the case when $N^\dagger\neq0$. In our model, it just corresponds to the case of $\beta\neq0$, while $\alpha=0$.

With this ansatz, it is found that the equations for $M^{(a)}_{tr}$ are algebraic ones  and can be solved directly~\cite{Cai:201412}. Therefore we pay main attention on $\alpha$ and $\beta$. At the horizon, the regular initial conditions should be  imposed. The behavior of the solutions of equations in the UV region (near the AdS boundary)  depends on the value of $m^2+k$. When $m^2+k=0$, the asymptotic solutions will have a logarithmic term, we will not consider this case here.  Instead when $m^2+k\neq0$, we have the asymptotic solution as~\cite{Cai:201412}
\begin{equation}\label{ab2}
\begin{split}
&\alpha_{UV}=\alpha_+r^{(1+\delta_1)/2}+\alpha_-r^{(1-\delta_1)/2}-\frac{B}{m^2+k},\\
&\beta_{UV}=\beta_+r^{(1+\delta_2)/2}+\beta_-r^{(1-\delta_2)/2},\\
&\delta_1=\sqrt{1+4k+4m^2},~~\delta_2=\sqrt{1-4k+4m^2},
\end{split}
\end{equation}
where $\alpha_{\pm}$ and $\beta_{\pm}$ are all finite constants. To make the system condense into the antiferromagnetic phase,  as in  Ref.~\cite{Cai:2014jta}, we require that the term associated with the magnetic field $B$  in Eq.~\eqref{ab2} should be the leading term.  When $B=0$, we require that the condensation for $\beta$ appears spontaneously. With those, the  parameters have to  satisfy $m^2>k>0$ and
\begin{equation}\label{canshu}
J_c^+(k,m^2)< J<J_c^-(k,m^2),
\end{equation}
with $J_c^\pm(k,m^2)=-(m^2+k)^2(m^2+3/2\pm k)/12$ and $\alpha_+=\beta_+=0$ according to  the AdS/CFT dictionary (for details please see \cite{Cai:201412}).

\begin{figure}
\includegraphics[width=0.22\textwidth]{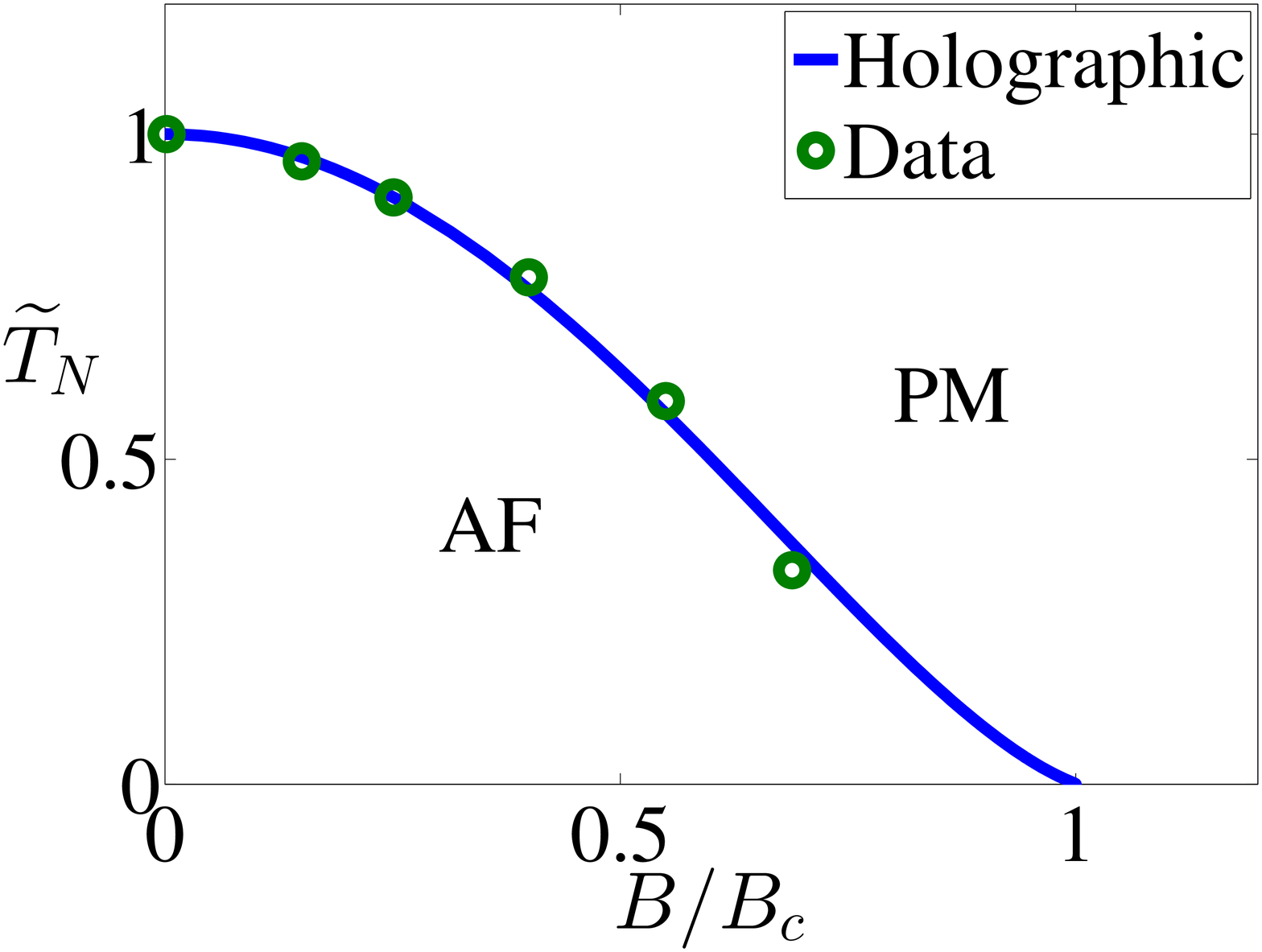}
\includegraphics[width=0.22\textwidth]{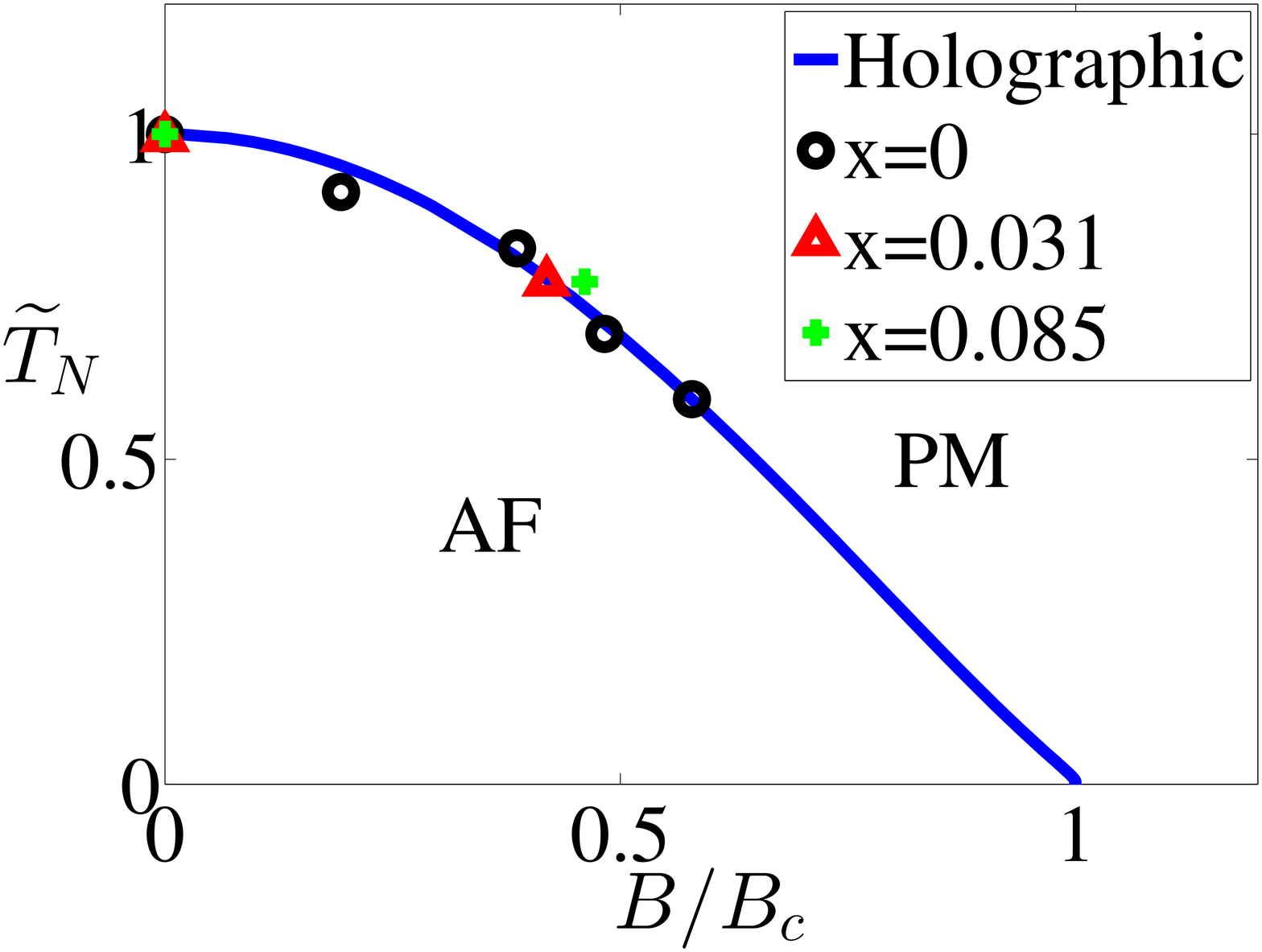}
\caption{The antiferromagnetic critical temperature $T_N$ versus the external magnetic field $B$  compared with experimental data. The critical temperature $T_N$ is calculated with the aid of the solution of  Eqs.~(\ref{eqab2}). At each fixed value of magnetic field $B$  the temperature is determined  at which the AF condensate starts to appear.
 {\bf Left}: Comparion with experimental data from  BiCoPO$_5$ {\bf Right}: Comparion with experimental data for pyrochlore compounds:  Er$_{2-2x}$Y$_{2x}$Ti$_2$O$_7$.
The experimental data are from ~\cite{Mathews-2013,Niven-2014} and rescaled.}
\label{TNB1}
\end{figure}

{\emph {QCP, energy gap and spectrum.}}--Let us first consider the influence of the external magnetic field $B$ on the antiferromagnetic critical temperature $T_N$. Near the critical temperature, the staggered magnetization is very small, i.e., $\beta$ is a small quantity. In that case we can neglect the nonlinear terms of $\beta$ and obtain the equations for $\alpha$ and $\beta$,
\begin{equation}\label{eqab2}
\alpha''+\frac{f'\alpha'}f-\frac{m^2_{\alpha\text{eff}}}{r^2f}\alpha=\frac{B}{r^2f}, ~\beta''+\frac{f'\beta'}f-\frac{m^2_{\beta\text{eff}}}{r^2f}\beta=0.
\end{equation}
Here $m^2_{\alpha\text{eff}}$ and $m^2_{\beta\text{eff}}$ are two functions of $\alpha$~\cite{Cai:201412}. Without loss of generality, we can set $\beta(r_h)=1$. With  increasing the magnetic field $B$ from zero, the effective mass square of $\beta$ increases, so that the critical temperature $T_N$ decreases.  The critical temperature is plotted as a function of the external magnetic field in  Fig.~\ref{TNB1}. When $T_N$ is decreased to zero, an AdS$_2$ geometry emerges near the horizon~\cite{Cai:201412}. The existence of a stable IR fixed point in the emergent AdS$_2$ region demands
\begin{equation}\label{expBc}
B=B_c\equiv-m^2_{\alpha\text{eff}}|_{\alpha=\alpha_c},~m^2_{\beta\text{eff}}|_{\alpha=\alpha_c}=0
\end{equation}
at the horizon $r=r_h=1$. Then we can see that in  the case of $T=0$, when $|B|<|B_c|$, $\beta$ is unstable near the horizon and the condensation  happens so that  the staggered magnetization is no longer
vanishing.  When $|B|>|B_c|$,  however, $\beta$ is stable at the horizon and the staggered magnetization is zero. Therefore, a QPT occurs at $|B|=|B_c|$ and the system is quantum disorder when $|B|>|B_c|$.

In order to investigate the magnetic fluctuations in the vicinity of QCP, we need to consider the perturbations of two polarization fields. To make the system be self-consistent at the linear order, the perturbations for all components of the polarization fields have to be considered,
\begin{equation}\label{pertbur1}
\begin{split}
&\delta M^{(a)}_{\mu\nu}=\epsilon C^{(a)}_{\mu\nu}e^{-i(\omega t+qx)},~~(\mu,\nu)\neq(r,y), (t,x)\\
&\delta M^{(a)}_{\mu\nu}=i\epsilon C^{(a)}_{\mu\nu}e^{-i(\omega t+qx)},~~(\mu,\nu)=(r,y), (t,x).
\end{split}
\end{equation}
Put this perturbations into the equation of motions and compute to the 1st order for $\epsilon$, we can get their equations of the perturbations (the details for perturbational equations can be found in Ref.~\cite{Cai:201412}). In general, because of the nonlinear potential, all the components of two polarization field couple with each other. Let $\widetilde{\beta}=(C^{(1)}_{xy}-C^{(2)}_{xy})/2$. In the paramagnetic magnetic phase ($T>T_N$ or $B>B_c$),  when $q, \omega\rightarrow0$, the equations for $\widetilde{\beta}$ decouple from others~\cite{Cai:201412}. By imposing the ingoing condition at the horizon, it has the following asymptotic solution in the UV region,
\begin{equation}\label{asy1}
\widetilde{\beta}\simeq\widetilde{\beta}_+r^{(1+\delta_2)/2}+\widetilde{\beta}_-r^{(1-\delta_2)/2}.
\end{equation}
According to the dictionary of AdS/CFT, up to a positive constant, the retarded Green's function for $\widetilde{\beta}$ reads
\begin{equation}\label{Green2}
G_{\beta\beta}=\widetilde{\beta}_-/\widetilde{\beta}_+.
\end{equation}
Using the retarded Green's function, we can define spectrum function as
$P(\omega,\overrightarrow{q})=\text{Im}\, G(\omega,\overrightarrow{q})/\pi$.
When we turn on a small momentum $\overrightarrow{q}$, the energy of long-life quasi-particle, which corresponds to the peak of $P(\omega, \overrightarrow{q})$, can be given by following dispersion relation,
\begin{equation}\label{dispre1}
    \omega_*=\Delta+\epsilon_{\overrightarrow{q}},~~\epsilon_{\overrightarrow{q}=0}=0.
\end{equation}
Here $\Delta$ is the energy gap of quasi-particle excitation. In the vicinity of QCP, for the case of $\omega=0$, the retarded Green's function usually has the form of $G\sim1/(q^2+1/\xi^2)$, where $\xi$ is called correlation length. At QCP, in general, the energy gap vanishes. Thus we have $\omega_*=\epsilon_{\overrightarrow{q}}$. In addition, for small frequency and wave vector, we can define the dynamic exponent $z$ in the way as $\omega_*\sim q^z$.

%

\begin{figure}
\includegraphics[width=0.22\textwidth]{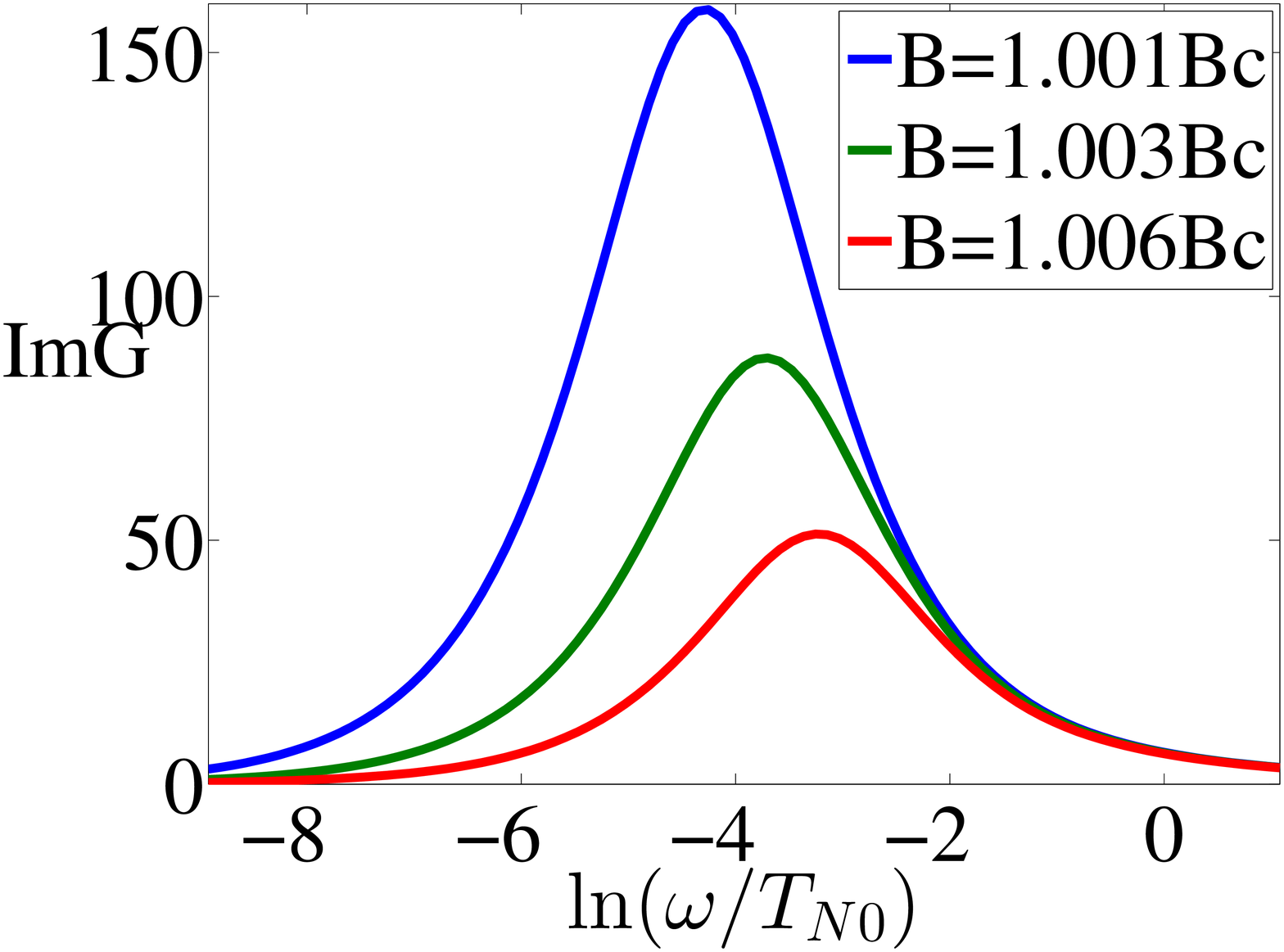}
\includegraphics[width=0.22\textwidth]{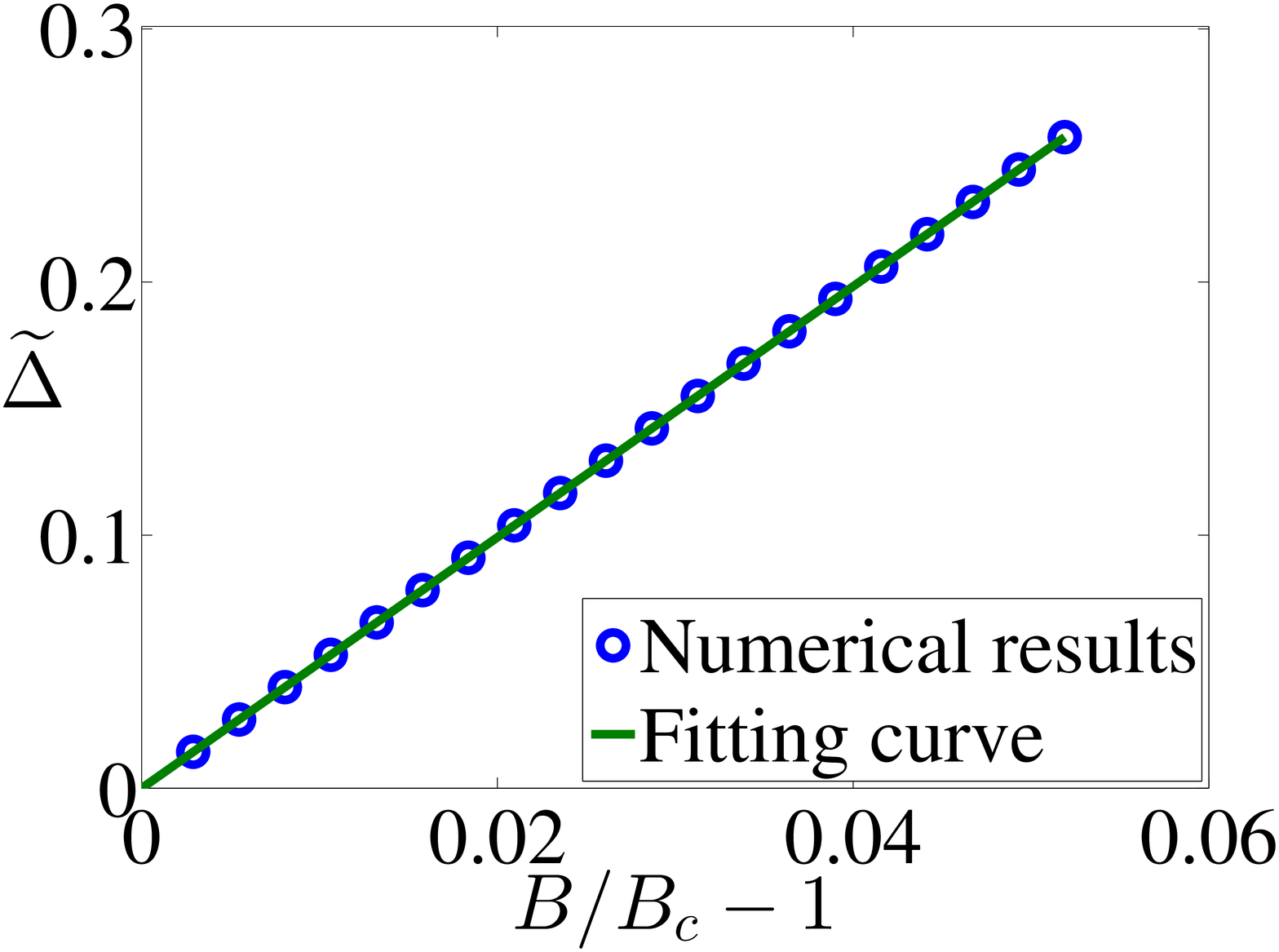}
\caption{{\bf Left}: The antiferromagnetic spectrum function in the case with different magnetic field when $B>B_c$. {\bf Right}: The  gap energy versus the external magnetic field when $B/B_c-1\rightarrow0^+$.}. 
\label{BG1}
\end{figure}

\begin{figure}
\includegraphics[width=0.22\textwidth]{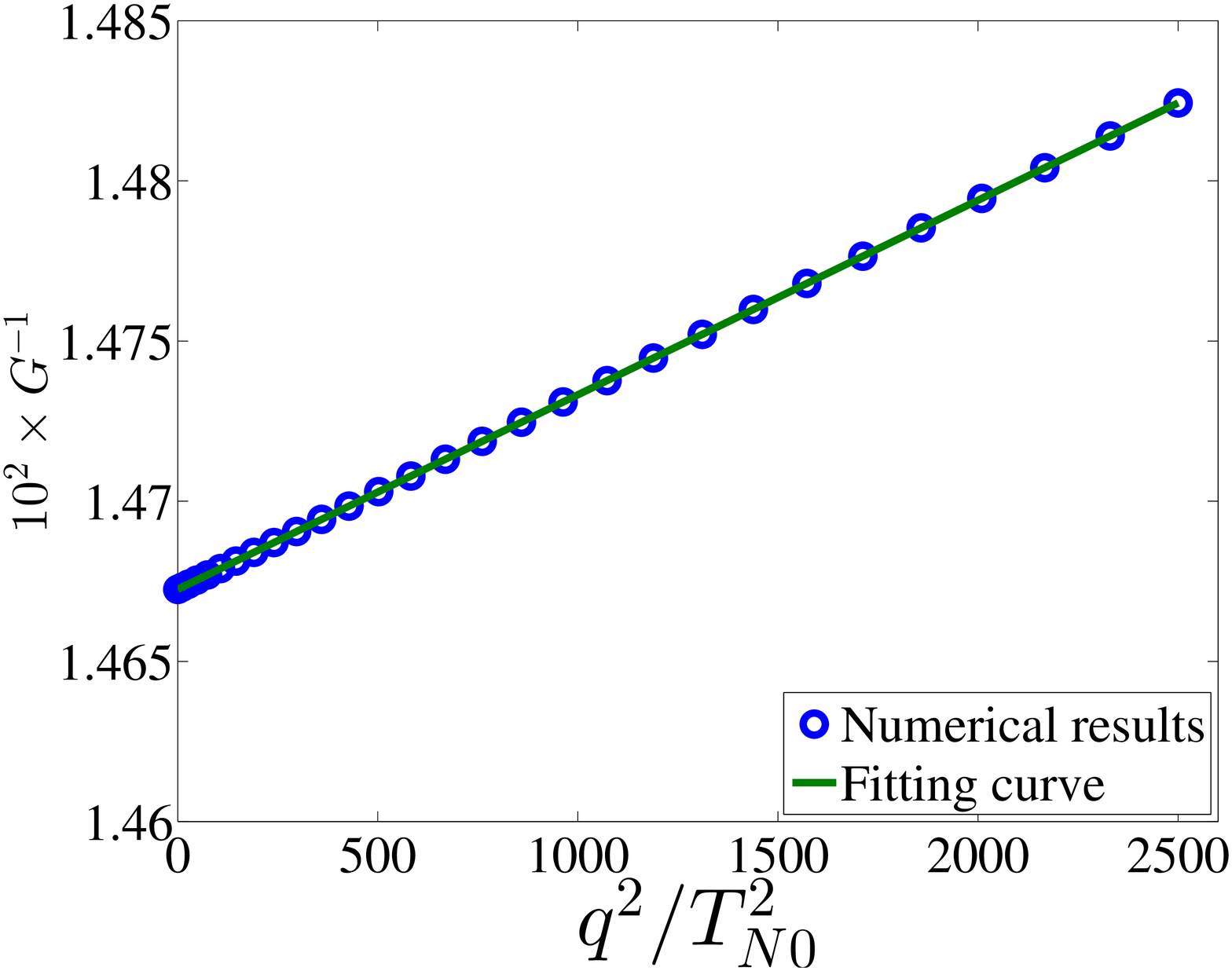}
\includegraphics[width=0.22\textwidth]{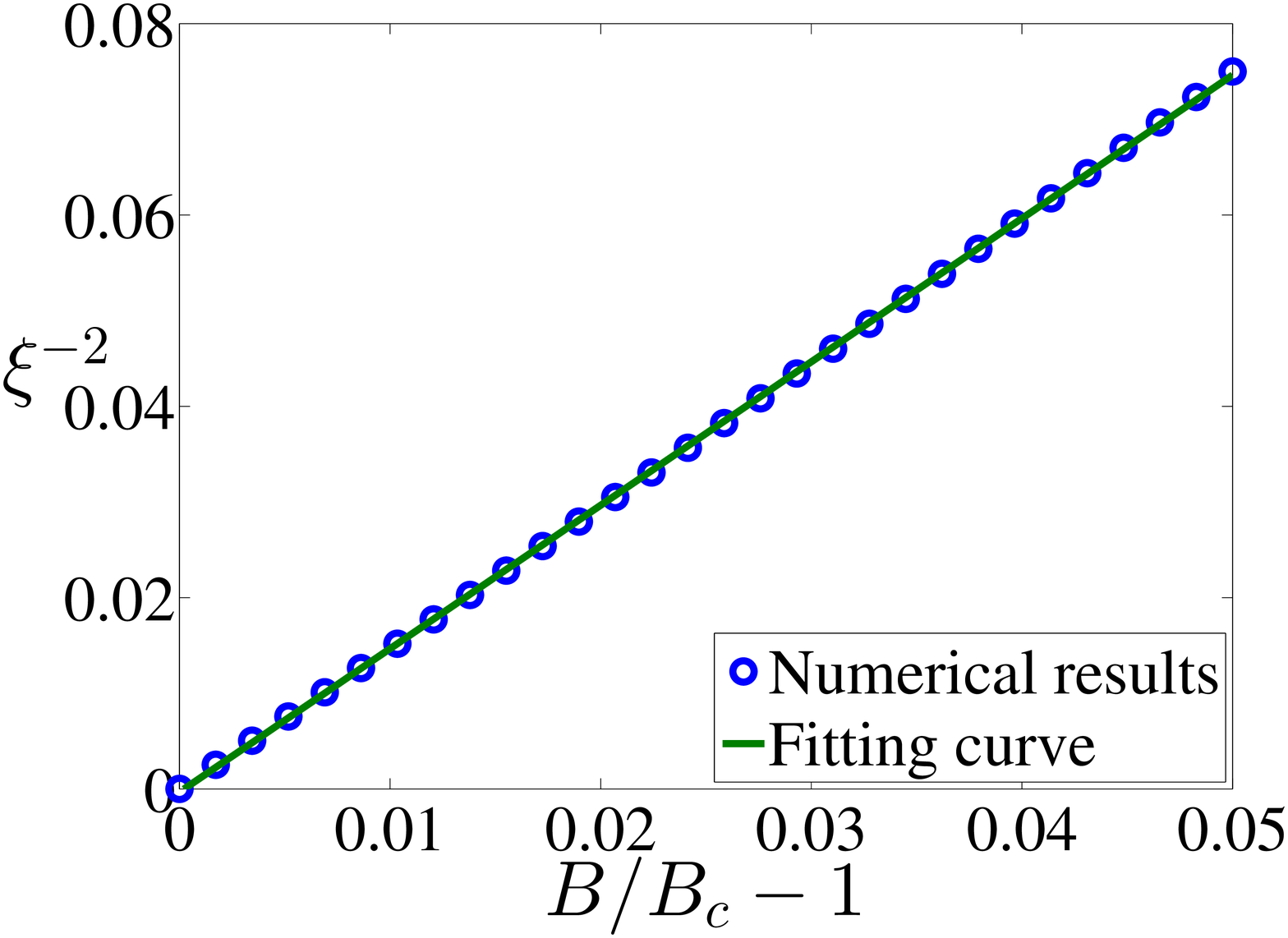}
\caption{{\bf Left}:  $G^{-1}$ as a function of $q^2$ when $\omega=0$ in the case of $|1-B/B_c|=0.01$. The solid line is the fitting curve by $G^{-1}\propto q^2+1/\xi^2$. {\bf Right}: The  correlation length $\xi$ versus the  magnetic field when $|B/B_c-1|\rightarrow0$.}. 
\label{BL1}
\end{figure}

\indent{\it Numerical Results.}
As the equations involved here are nonlinear, we have to solve them numerically. The different parameters satisfying restrictions~\eqref{canshu} give similar results, we here therefore just take parameters as  $m^2=1,k=7/8$ and $J=-0.67$ as a typical example in the  left plot of Fig.~\ref{TNB1},  and Figs. \ref{BG1} and~\ref{BL1} and  as $J=-0.71$ in the right plot of Fig.~\ref{TNB1}

We can see from Fig.~\ref{TNB1} that the N\'{e}el temperature $T_N$ is suppressed by external magnetic field. There is a critical magnetic field for given parameters, at which $T_N$ is zero and QPT occurs. After rescaled, different parameters give similar behaviors with some slight  differences. The physical picture of this QPT can be understood as follows. When magnetic field reaches its critical value, there magnetic spins become partially aligned along the direction of the magnetic field.  Therefore the system requires less thermal energy to destroy the remaining magnetic spins order.

The holographic model can  give some interesting scaling relations near the QCP. For small $B$, numerical results show that $T_N-T_{N0}\propto B^2$, where $T_{N0}$ denotes the critical temperature
in the case without external magnetic field. When magnetic field is close to $B_c$, we find that N\'{e}el temperature is fitted well by following relation
\begin{equation}\label{TB1}
\widetilde{T}_N/\ln \widetilde{T}_N\propto(1-B/B_c),
\end{equation}
where $\widetilde{T}_N=T_N/T_{N0}$. We will analytically present the relation (\ref{TB1})  by considering the emergent geometry $AdS_2$  in the  IR limit~\cite{Cai:201412}.

When magnetic field $B$ is larger than the critical value $B_c$, the antiferromagnetic phase disappears  even at zero temperature. In this case, the system comes into quantum disordered phase at zero temperature, in which there is a gapped magnetic excitation. In the left plot of Fig.~\ref{BG1}, we show $\text{Im}\,G$ with respect to the frequency of antiferromagnetic excitation in the case with different magnetic field.
In the case of $0<B/B_c-1\ll1$, there is a distinct peak which gives the energy gap for the excitation. With increasing magnetic field, the peak moves towards higher energy and becomes more and more indistinct. This means that the gap increases but the lifetime decreases when magnetic field increases. At the critical magnetic field $B=B_c$, we see $\omega_*=0$, which corresponds to a gapless long-lifetime antiferromagnetic  excitation. In the region of $B/B_c-1\rightarrow0^+$, we find the energy gap is fitted well by following equation (see the right plot of Fig.~\ref{BG1})
\begin{equation}\label{TB2}
\widetilde{\Delta}\propto(B/B_c-1), \text{with}~\widetilde{\Delta}=\Delta/T_{N0}.
\end{equation}
In the left plot of Fig.~\ref{BL1}, we plot the inverse Green's function $G_{\beta\beta}^{-1}(q)$ in the case of $\omega=0$ and $|1-B/B_c|=0.01$. We can see that it obeys the behavior of $G^{-1}\sim q^2+1/\xi^2$ as we expected  before. Thus the Green's function can give the correlation length by fitting the curve of $G^{-1}$ as a function of $q^2$, which is shown in the right plot of Fig.~\ref{BL1}. We see that the correlation length $\xi$ as a function of the tuning parameter $B$ obeys the following relation
\begin{equation}\label{Bl0}
  \xi\propto|B/B_c-1|^{-\nu},~~\text{with}~\nu\simeq1/2.
\end{equation}
As to the dynamical exponent $z$,  in antiferromagnetic metals, $z=2$~\cite{Lohneysen.07}.   In our holographic model, the dynamical exponent can be calculated  by using similar numeric method.  The results indeed show that $z\simeq2$. This numerical result can be confirmed  by the emergent $AdS_2$ geometry in the IR region~\cite{Cai:201412} and agrees with the result from a different holographic model proposed in Ref.~\cite{Iqbal:2010eh}.  Furthermore, from the  energy gap (\ref{TB2}),  we see that this energy gap satisfies  the universal scaling relation $\Delta\sim|B-B_c|^{z\nu}$, which further indicates $z=2$ in this model.
%
%
%

\indent { \it Discussion.}
-The relation we found in this paper of the N\'{e}el temperature with respect to the magnetic field in the vicinity of QCP is quite non-trivial.  Note that the relation~\eqref{TB1} is not the usual power-law behavior or square-root form. But it is an expected result  in the 2-D QPTs in strong coupling case~\cite{Lohneysen.07}. This non-trivial coincidence strongly implies a connection between these two different theories. In the antiferromagnetic metal where magnetic ordered is dominated by itinerant electrons, dynamical exponent $z$ is 2 near the QCP. Since our dual boundary theory is  a 2-D theory, the effective dimension is thus $d_{\text{eff}}=d+z=4$, which is just the upper critical dimension of the Hertz field theory~\cite{Hertz.76,Ar.Abanov}. In this case, the hyperscaling is violated and logarithmic correction behavior appears. In fact, the $d=z=2$ quantum critical theory is in general not a weak coupling theory at any $T>0$. Instead, a strongly coupled effective classical model emerges that can be used to determine the critical dynamics~\cite{S.Sachdev2}.  Our results show that it can be described well by AdS/CFT correspondence and this provides a new example of the applicability of the gravity/gauge duality in condensed matter theory.


It is quite interesting to apply this holographic model to realistic materials.   Since the holographic model is independent of the microscopic details of the materials and their interactions,  it should be suitable for a class of materials. Two potential AF-QPT materials  are BiCoPO$_5$ with critical magnetic field $B_c\simeq15.3$T (which is obtained by fitting a power-law relation~\cite{Mathews-2013}) and  Er$_{2-2x}$Y$_{2x}$Ti$_2$O$_7$ with $B_c\simeq1.5$T for $x=0$~\cite{J.P.C.Ruff,Niven-2014}. In Fig. \ref{TNB1} we present the experimental data for these two materials and the holographic results, where we choose the model parameters so that they can give out the best fitting. The  holographic model gives  $B_c\simeq16.2$T and 1.45T, respectively. While the experiment data show that the energy gap for Er$_{2-2x}$Y$_{2x}$Ti$_2$O$_7$ when $B\geq B_c$ obeys the linear relationship~\eqref{TB2} with slope 4.2  (see  Fig.8 in Ref.~\cite{Niven-2014}), our holographic model  gives  the slope
 5.0  with the chosen model parameters. These two slopes are in the same order. 
 Note that these two kinds of material have different microscopic structures and complex interactions, it is remarkable that the simple holographic model can give a self-consistent description for those two
 materials.  In addition, it is also worth mentioning here  that the doping at the magnetic site ($x$ up to 0.085) has a very little influence on the critical behavior of Er$_{2-2x}$Y$_{2x}$Ti$_2$O$_7$. This indicates that an emergent universal behavior appears in these materials from very different microscopic details and could be described by the holographic model.

As the critical behavior of a QPT induced by a magnetic field, the three scaling relations \eqref{TB1}, \eqref{TB2} and \eqref{Bl0}  near the critical point are our main results from the holographic model.
Besides the energy gap (\ref{TB2}),  our predications on the scaling relations \eqref{TB1} and \eqref{Bl0}  can also be confirmed  by experiments. Unfortunately at the moment they cannot be checked by the existing experiments because the experiment data are absent when the N\'{e}el temperature is very close to zero. Of course  whether  the holographic model  is suitable for these two materials needs more evidence. It is also very interesting to find more  materials satisfying the conditions of this model and to check our predictions. We expect that this model can be confirmed or  falsified experimentally soon.

\section*{Acknowledgements}
We thank the quite helpful discussions with  J. Erdmenger and K. Schalm during the KITPC program ``Holographic duality for condensed matter physics"  (July 6-31, 2015, Beijing).
This work was supported in part by the National Natural Science Foundation of China (No.10821504, No.11035008, No.11375247, and No.11435006 ).




\begin{thebibliography}{99}


\bibitem{S. Sachdev}
S. Sachdev and B. Keimer,
``Quantum criticality",
Phys. Today {\bf 64} (2), 29 (2011).

\bibitem{Hertz.76}
J.~A.~Hertz,
``Quantum critical phenomena",
Phys.~Rev.~B {\bf 14}, 1165--1184 (1976).


\bibitem{Sachdev_book}
S.~Sachdev, \emph{Quantum Phase Transitions}, Cambridge University Press,
Cambridge (1999).



\bibitem{Gegenwart.08}
P.~Gegenwart, Q.~Si, and F.~Steglich,
``Quantum criticality in heavy-fermion metals",
Nat.~Phys. \textbf{4}, 186--197 (2008).

\bibitem{Lohneysen.07}
  H.~v.~Lohneysen, A.~Rosch, M.~Vojta and P.~Wolfle,
  ``Fermi-liquid instabilities at magnetic quantum phase transitions,''
  Rev.\ Mod.\ Phys.\  {\bf 79}, 1015 (2007).
\bibitem{G.Chaboussant}
G. Chaboussant, P.A. Crowell, L.P. Levy, O. Piovesana, A. Madouri,and D. Mailly,
``Experimental phase diagram of Cu2(C5H12N2)2Cl4: A quasi-one-dimensional antiferromagnetic spin-Heisenberg ladder",
Phys. Rev. B {\bf 55}, 3046 (1997).

\bibitem{W.Shiramura}
H. Tanaka, W. Shiramura, T. Takatsu, B. Kurniawan, M. Takahashi, K. Kamishima, K. Takizawa, H. Mitamura and T. Goto,
``High-Field Magnetization Processes of Double Spin Chain Systems KCuCl3 and TlCuCl3",
J. Phys. Soc. Jpn. {\bf 66}, 1900 (1997);

\bibitem{Oosawa}
A. Oosawa, T. Takamasu, K. Tatani, H. Abe, N. Tsujii, O. Suzuki, H. Tanaka, G. Kido, K. Kindo,
``Field-induced magnetic ordering in the quantum spin system KCuCl3",
Phys. Rev. B {\bf 66}, 104405 (2002).

\bibitem{Ch.03}
Ch. R\"{u}egg, N. Cavadini, A. Furrer, H.-U. G\"{e}del, K. Kr\"{a}mer, H. Mutka, A. Wildes, K. Habich and P. Vorderwisch,
``Bose¨CEinstein condensation of the triplet states in the magnetic insulator TlCuCl3",
Nature {\bf 423}, 62 (2003);

\bibitem{Nikuni}
T. Nikuni, M. Oshikawa, A. Oosawa, and H. Tanaka,,
``Bose-Einstein Condensation of Dilute Magnons in TlCuCl3",
Phys. Rev. Lett. {\bf 84} 5868 (2000).

\bibitem{Si.01}
Q.~Si, S.~Rabello, K.~Ingersent, and J.~L.~Smith,
``Locally critical quantum phase transitions in strongly correlated metals",
Nature \textbf{413}, 804--808 (2001).

\bibitem{Senthil.04}
T.~Senthil, A.~Vishwanath, L.~Balents, S.~Sachdev, and M.~P.~A. Fisher,
``Deconfined Quantum Critical Points",
Science \textbf{303}, 1490--1494 (2004);

\bibitem{Tanaka}
H. Tanaka, A. Oosawa, T. Kato, H. Uekusa, Y. Ohashi, K. Kakurai, and A. Hoser,
``Observation of Field-Induced Transverse N\'{e}el Ordering in the Spin Gap System TlCuCl3",
J. Phys. Soc. Jpn. {\bf 70}, 939 (2001).



\bibitem{Maldacena:1997re}
 J.~M.~Maldacena,
  ``The large N limit of superconformal field theories and supergravity,''
  Adv.\ Theor.\ Math.\ Phys.\  {\bf 2}, 231 (1998)
  [Int.\ J.\ Theor.\ Phys.\  {\bf 38}, 1113 (1999)]

\bibitem{Gubser:1998bc}
  S.~S.~Gubser, I.~R.~Klebanov and A.~M.~Polyakov,
  ``Gauge theory correlators from non-critical string theory,''
 Phys.\ Lett.\  B {\bf 428}, 105 (1998)



\bibitem{Witten:1998qj}
  E.~Witten,
  ``Anti-de Sitter space and holography,''
  Adv.\ Theor.\ Math.\ Phys.\  {\bf 2}, 253 (1998)

\bibitem{Hartnoll:2008vx}
  S.~A.~Hartnoll, C.~P.~Herzog and G.~T.~Horowitz,
  ``Building a Holographic Superconductor,''
  Phys.\ Rev.\ Lett.\  {\bf 101}, 031601 (2008)

\bibitem{Lee:2008xf}
  S.~S.~Lee,
  ``A Non-Fermi Liquid from a Charged Black Hole: A Critical Fermi Ball,''
  Phys.\ Rev.\ D {\bf 79}, 086006 (2009)

\bibitem{Liu:2009dm}
  H.~Liu, J.~McGreevy and D.~Vegh,
  ``Non-Fermi liquids from holography,''
  Phys.\ Rev.\ D {\bf 83}, 065029 (2011)

\bibitem{Cubrovic:2009ye}
  M.~Cubrovic, J.~Zaanen and K.~Schalm,
  ``String Theory, Quantum Phase Transitions and the Emergent Fermi-Liquid,''
  Science {\bf 325}, 439 (2009)

\bibitem{Cai:2014oca}
  R.~-G.~Cai and R.~-Q.~Yang,
  ``Paramagnetism-Ferromagnetism Phase Transition in a Dyonic Black Hole,''
  Phys. Rev. D {\bf 90}, 081901 (2014)

\bibitem{Cai:2014jta}
R.~G.~Cai and R.~Q.~Yang,
  ``Holographic model for the paramagnetism/antiferromagnetism phase transition,''
  Phys.\ Rev.\ D {\bf 91}, no. 8, 086001 (2015)

\bibitem{Cai:2015bsa}
  R.~G.~Cai and R.~Q.~Yang,
  ``Antisymmetric tensor field and spontaneous magnetization in holographic duality,''
  [arXiv:1504.00855 [hep-th]].


\bibitem{Cai:1996eg}
  R.~-G.~Cai and Y.~-Z.~Zhang,
  ``Black plane solutions in four-dimensional space-times,''
  Phys.\ Rev.\ D {\bf 54}, 4891 (1996)  

\bibitem{Cai:201412}
  R.~G.~Cai, R.~Q.~Yang and F.~V.~Kusmartsev,
  ``Holographic antiferromganetic quantum criticality and AdS$_2$ scaling limit,''
  [arXiv:1505.03405 [hep-th]].


\bibitem{Mathews-2013}
 E. Mathews, K. M. Ranjith, M. Baenitz and R. Nath,
``Field induced magnetic transition in low-dimensional magnets Bi(Ni,Co)PO5",
Solid State Communications {\bf 154} 56-59 (2013) .

\bibitem{J.P.C.Ruff}
J.P.C. Ruff, J.P. Clancy, A. Bourque, M.A. White, M. Ramazanoglu, J.S. Gardner, Y. Qiu, J.R.D. Copley, M.B. Johnson, H.A. Dabkowska and B.D. Gaulin,
``Spin Waves and Quantum Criticality in the Frustrated XY Pyrochlore Antiferromagnet Er$_2$Ti$_2$O$_7$",
Phys. Rev. Lett. {\bf 101} 147205 (2008)


\bibitem{Niven-2014}
J.F. Niven, M. B Johnson ,A. Bourque, P.J. Murray, D. D. James, H. A. Dabkowska , B. D. Gaulin and M. A. White,
``Magnetic phase transitions and magnetic entropy in the XY antiferromagnetic pyrochlores(Er$_{1-x}$Y$_x$ )$_2$Ti$_2$O$_7$".
Proc. R. Soc. A {\bf 470}: 20140387 (2014). 

\bibitem{Iqbal:2010eh}
  N.~Iqbal, H.~Liu, M.~Mezei and Q.~Si,
  ``Quantum phase transitions in holographic models of magnetism and superconductors,''
  Phys.\ Rev.\ D {\bf 82}, 045002 (2010)  

\bibitem{Ar.Abanov}
Ar. Abanov, A. Chubukov,
``Anomalous Scaling at the Quantum Critical Point in Itinerant Antiferromagnets",
Phys.~Rev.~Lett. {\bf 93} 255702 (2004);

\bibitem{S.Sachdev2}
 S. Sachdev, and E. R. Dunkel,
 ``Quantum critical dynamics of the two-dimensional Bose gas",
 Phys. Rev. B {\bf 73}, 085116 (2006)




\end{thebibliography}
\end{document}